\documentclass[a4paper,10pt]{article}
\pdfoutput=1 
\usepackage{jheppub}
\usepackage[T1]{fontenc} 
\usepackage{float}
\usepackage{tikz-cd}
\usepackage{braket}
\usetikzlibrary{cd}
\definecolor{rindou1}{rgb}{0.4431,0.2862,0.7960}
\definecolor{rindou2}{rgb}{0.0078,0.1215,0.4392}
\definecolor{lapis}{rgb}{0.0.0470,0.2941,0.5568}
\definecolor{emerald}{rgb}{0.31, 0.78, 0.47}
\definecolor{deepsaffron}{rgb}{1.0, 0.6, 0.2}
\definecolor{pinegreen}{rgb}{0.0, 0.47, 0.44}
\definecolor{majorelleblue}{rgb}{0.38, 0.31, 0.86}
\definecolor{jade}{rgb}{0.0, 0.66, 0.42}
\definecolor{teal}{rgb}{0.0, 0.5, 0.5}
\definecolor{darkcyan}{rgb}{0.0, 0.55, 0.55}
\definecolor{jazzberryjam}{rgb}{0.65, 0.04, 0.37}
\definecolor{electricviolet}{rgb}{0.56, 0.0, 1.0}
\definecolor{regalia}{rgb}{0.32, 0.18, 0.5}
\definecolor{burgundy}{rgb}{0.5, 0.0, 0.13}
\definecolor{indigo(web)}{rgb}{0.29, 0.0, 0.51}
\definecolor{cerise}{rgb}{0.87, 0.19, 0.39}
\definecolor{darkbyzantium}{rgb}{0.36, 0.22, 0.33}
\definecolor{darkscarlet}{rgb}{0.34, 0.01, 0.1}
\definecolor{tyrianpurple}{rgb}{0.4, 0.01, 0.24}
\definecolor{bondiblue}{rgb}{0.0, 0.58, 0.71}
\definecolor{amaranth}{rgb}{0.9, 0.17, 0.31}
\definecolor{blue(munsell)}{rgb}{0.0, 0.5, 0.69}
\definecolor{yaleblue}{rgb}{0.06, 0.3, 0.57}
\definecolor{iris}{rgb}{0.35, 0.31, 0.81}
\definecolor{darkred}{rgb}{0.55, 0.0, 0.0}
\definecolor{slateblue}{rgb}{0.42, 0.35, 0.8}
\usepackage[T1]{fontenc}
\usepackage{datetime2}
\usepackage[english]{babel}
\usepackage[utf8]{inputenc}
\usepackage{graphicx}
\usepackage{color}
\usepackage{amsfonts,amsthm}
\usepackage{bm,bbm}
\usepackage{mathrsfs}
\usepackage{amsmath}
\usepackage{float}
\usepackage{pgf,tikz}
\usepackage{mathrsfs}
\usepackage{xcolor}
\usetikzlibrary{arrows}

\newcommand{\z}{\overline{z}}

\title{\boldmath Soft Gravity by Squaring Soft QED on the Celestial Sphere}

\author{Nikhil Kalyanapuram}

\affiliation{Department of Physics and Institute for Gravitation and the Cosmos, The Pennsylvania State University, University Park PA 16802, USA}

\emailAdd{nkalyanapuram@psu.edu}

\abstract{We recast the soft $S$-matrices on the celestial sphere as correlation functions of certain $2$-dimensional models of topological defects. In pointing out the double copy structure between the soft photon and soft graviton cases, we arrive at a putative classical double copy between the corresponding topological models and a rederivation of gauge invariance and the equivalence principle as Ward identities of the $2$-dimensional theories.}

\begin{document} 
\maketitle
\flushbottom

\section{Introduction}\label{sec:intro}
The study of scattering amplitudes as analytical objects in their own right has been a rather fertile one in recent years. Of particular interest has been the soft sector of the $S$-matrix for theories with massless particles, which as it turns out reflects underlying symmetries which are manifest at null infinity (see \cite{Strominger:2015bla,Kapec:2015ena,He:2015zea,Strominger:2014pwa,He:2014cra,He:2014laa,He:2017fsb,A:2020lub,Campiglia:2019wxe,Campiglia:2018dyi,Laddha:2017vfh,Campiglia:2016efb,Campiglia:2016hvg,Campiglia:2015kxa,Campiglia:2015qka,Campiglia:2014yka,R:2018fup,Strominger:2017zoo,Banerjee:2019prz,Banerjee:2020kaa,Banerjee:2020vnt,Banerjee:2020zlg} and references therein). 

Emerging from this analysis is a picture of flat space holography, in which (massless) particles on the celestial sphere control most of the analytic properties of the entire $S$-matrix. The representation of the $S$-matrix in terms of operators on the celestial sphere is realized by a change of basis brought about by the \emph{Mellin transform} \cite{Pasterski:2016qvg,Pasterski:2017ylz}. In this basis, the infrared sector of the $S$-matrix can be understood with relative ease in terms of operator product expansions (OPEs) of insertions on $\mathbb{CP}^{1}$ \cite{Nande:2017dba,Pate:2019mfs,Pate:2019lpp,Himwich:2020rro}. This has led to a conjectural $4d/2d$ duality, which relates the $S$-matrix to correlation functions of operators of an alleged CFT on the celestial sphere.

In this short note, in focusing our attention on the soft $S$-matrix, we will realize an amusing web of dualities. The soft $S$-matrix for photons is recast as a correlation function of a Coulomb gas, which is a manifestly two-dimensional system. A double copy construction relating the soft photon $S$-matrix to the one in gravity is manifested by squaring the kinetic operator of the dual $2d$ model. The resulting model in two dimensions is identified with a dynamical system of crystal disclinations. These observations lead to a rather satisfying web of dualities which to our knowledge has been stated in this manner for the first time. The duality conveniently realizes the soft theorems as a consequence of shift symmetries in the two-dimensional models, supplying the theorems of charge conservation and the equivalence principle of Weinberg \cite{Weinberg:1964ew} as manifestations of the corresponding Ward identities.

\paragraph{Note:}
Although we have in many instances used the word duality, it is meant in an unconventional sense. While dualities are normally meant to express equivalences between strongly and weakly coupled theories, the dualities we discuss are all between weakly coupled theories.

\section{A Duality on the Celestial Sphere}
Suppose we have a scattering process with $n$ massless charged particles. For purposes of simplicity, we assume that all particles are outgoing. We focus our attention on that part of the \emph{infrared divergent} $S$-matrix receiving contributions from purely virtual photons, defined according to the factorization theorem \cite{Collins:1989bt,Gardi:2009qi}

\begin{equation}
    \mathcal{A}_{m,s=1} = \mathcal{C}_{n}\times\mathcal{A}^{soft}_{n,s=1}|_{vir}\times \mathcal{A}^{soft}_{n,s=1}|_{real}\times \mathcal{H}_{n,s=1}
\end{equation}
In this, the function $\mathcal{A}_{n,s=1}$ is the scattering amplitude for $n$ charged particles. $\mathcal{C}_{n}$ isolates singularities due to collinear sectors (we have compressed the notation somewhat - $\mathcal{C}_{n}$ itself is resolved into jet functions, one for each collinear sector). $\mathcal{H}_{n,s=1}$ is a hard function computed as a Wilson coefficient. Recently it has transpired that it may admit of an intrinsic definition \cite{Hannesdottir:2019opa}. $\mathcal{A}_{n,s=1}^{soft}|_{real}$ encodes soft singularities due to real emissions. The term of interest for us however is the term $\mathcal{A}_{n,s=1}^{soft}|_{vir}$, which represents virtual divergences due to virtual soft transmissions.

For the leading order soft theorem, this portion of the $S$-matrix is especially simple, and can be written in dimensional regularization\footnote{The dimensional regularization scheme used here makes application of the fact that IR divergences can be controlled beyond $4$ dimensions.} ($d = 4+\epsilon$) as \cite{Weinberg:1965nx}

\begin{equation}
    \ln\left(\mathcal{A}^{soft}_{n,s=1}|_{vir}\right) = -\frac{1}{8\pi^{2}\epsilon}\sum_{i\neq j}e_{i}e_{j}\ln|z_{i}-z_{j}|^{2}
\end{equation}
which is exact with the assumption that the theory has no interacting light matter.
For the case of gravity, the factorization theorem to leading order is far simpler, and takes the form

\begin{equation}
  \mathcal{A}_{m,s=2} = \mathcal{A}^{soft}_{n,s=2}|_{vir}\times \mathcal{A}^{soft}_{n,s=2}|_{real}\times \mathcal{H}_{n,s=2}  
\end{equation}
in accordance with the fact that there are no collinear divergences in gravity at leading order \cite{Weinberg:1965nx,Akhoury:2011kq}. We'll treat the case of gravity in the next section.

The $z_{i}$ are points on $\mathbb{CP}^{1}$, determined by directions of the external momenta according to the prescription\footnote{A definition more familiar to readers conversant with soft-collinear theory is obtained by writing $z_{i} = p^{\pm}_{i} = p^{x}_{i}\pm i p^{y}_{i}$.}
\begin{equation}
    p_{k} = \omega_{k}\left(1+z_{k}\z_{k},z_{k}+ \z_{k},-i(z_{k}-\z_{k}),1-z_{k}\z_{k}\right).
\end{equation}

We point out that the divergence manifested here cancels in suitably inclusive processes \cite{PhysRev.52.54,Kinoshita:1962ur,Lee:1964is,Weinberg:1965nx}. It has been established that this object can be represented as correlation functions of Wilson lines defined using asymptotic values of the gauge field and metric for the spin-$1$ \cite{Nande:2017dba} and spin-$2$ cases \cite{Himwich:2020rro} respectively\footnote{Wilson lines can also be used to extract the soft \emph{theorems} as well, and were applied in proofs of soft-collinear factorization in the past - see \cite{Collins:1988ig,Collins:1989gx,Feige:2013zla,Feige:2014wja,Feige:2015rea} and references therein.}.

Moving now to the first leg of the dualities, let the following action be noted

\begin{equation}\label{eq:3}
    S_{1} = 8\pi\epsilon\int d^{2}z \varphi(z,\overline{z})\partial\overline{\partial}\varphi(z,\overline{z}).
\end{equation}
The $\epsilon$ is the dimensional regularisation parameter $d-4$. This is easily identified as the Coulomb gas CFT in stereographic coordinates. It exhibits the well known BKT transition due to vortex binding below a critical temperature\footnote{The BKT transition, named after Brezenskii, Kosterlitz and Thouless is a topological phase transition. Unlike conventional Landau transitions, it is precipitated by the formation of topologically nontrivial excitations, which condense below a critical temperature. See \cite{Jha:2020oik} for an overview of numerical estimates of the BKT transition temperature.}.

The two point function is simply the Green's function on $\mathbb{CP}^{1}$

\begin{equation}
    \langle{\varphi(z_{1},\overline{z}_{1}),\varphi(z_{2},\overline{z}_{2})\rangle} = \frac{1}{8\pi^2\epsilon}\ln|z_{1}-z_{2}|^{2}.
\end{equation}
We find by direct computation that the vertex operators defined as

\begin{equation}
    V_{j}(z_{j},\overline{z}_{j}) = :e^{i e_{j}\varphi(z_{j},\overline{z}_{j})}:
\end{equation}
lead to the equivalence

\begin{equation}\label{eq:6}
    \langle{V_{1}(z_{1},\overline{z}_{1})\cdots V_{n}(z_{n},\overline{z}_{n})\rangle} = A_{n}^{soft}|_{vir,s=1}.
\end{equation}
We should like to emphasize that this is not a correlator of Wilson lines - it computes the correlation function of vertex operators in the Coulomb gas theory. In other words, the soft $S$-matrix can be derived equivalently as a correlator of vertex operators belonging to a dual two-dimensional model of interacting electrons.

So far the relationship discussed is purely formal and not especially surprising; the form of the integrated soft contributions lends itself quite easily to this dual picture. However, we would now like to understand how this observation can help inform a new perspective into the double copy on the celestial sphere\footnote{See \cite{Casali:2020vuy} for a Mellin space perspective at low multiplicity.}. The following section is dedicated to laying out the argument that realizes this hope. To do so, we turn our attention to the soft $S$-matrix for gravitationally interacting particles.

\section{A Double Copy on the Celestial Sphere}
The soft part of the $S$-matrix can be defined for any theory with massless quanta. In this section, we are concerned with the case of interacting spin-$2$ quanta. Accordingly, the virtual particles circulating in soft loops are gravitons. The soft $S$-matrix for such theories takes the form

\begin{equation}
\begin{aligned}
     &\ln\left(\mathcal{A}^{soft}_{n, s=2}|_{vir}\right) =\\ &-\frac{1}{8\pi^{2}\epsilon}\sum_{i\neq j}\kappa_{i}\kappa_{j}\omega_{i}\omega_{j}|z_{i}-z_{j}|^{2}\ln|z_{i}-z_{j}|^{2}
\end{aligned}
\end{equation}
where $\kappa_{i}$ is the coupling constant characterizing the strength of interaction between the corresponding particle and the graviton. We know however that the $\kappa_{i}$ must be the same for all $i$ by the equivalence principle. It will be seen that this need not be assumed; it arises as an implication of Ward identities of the two-dimensional model to be described.

It is worthwhile asking if there exists a double copy construction that relates the soft $S$-matrices of spin-$1$ and spin-$2$ transitions. Nominally, double copy statements like the ones due to Kawai, Lewellen and Tye (KLT) \cite{Kawai:1985xq} and to Bern, Carrasco and Johansson (BCJ) \cite{Bern:2019prr} rely on string-like descriptions \cite{DHoker:1989cxq,Mizera:2019gea,Mizera:2019blq,Kalyanapuram:2021xow,Kalyanapuram:2021vjt} or Feynman expansions. In this section, we show that this question can be answered directly at the level of two-dimensional dual models.

Take the kernel

\begin{equation}\label{eq:7}
   K(z_{ij}) = |z_{ij}|^{2}\ln|z_{ij}|^{2}
\end{equation}
where $z_{ij} = z_{i}-z_{j}$. Now, by direct calculation we have the result,

\begin{equation}
    \partial_{z_{i}}\overline{\partial}_{z_{i}}K(z_{ij}) \sim \ln|z_{ij}|^{2}
\end{equation}
where the $\sim$ is used to indicate equality up to an additive constant and we have chosen to take the derivative with respect to $z_{i}$ as a matter of convention. From this we have

\begin{equation}
    (\partial\overline{\partial})^{2}K(z) = \delta^{2}(z,\overline{z})
\end{equation}
which indicates that the kernel $K$ is the Green's function of the square of the Laplacian $\Delta=\partial\overline{\partial}$. In light with this, suppose we have a nonlinear sigma model defined by the action,

\begin{equation}\label{eq:3.5}
    S_{2} = 8\pi\epsilon\int d^{2}z c(z,\overline{z})(\partial\overline{\partial})^{2}c(z,\overline{z}).
\end{equation}
With the vertex operators

\begin{equation}
    U_{j}(z_{j},\overline{z}_{j}) = :e^{i\kappa_{j}\omega_{j}c(z_{j},\overline{z}_{j})}:
\end{equation}
it is the case that

\begin{equation}\label{eq:13}
    \mathcal{A}^{soft}_{n, s=2}|_{vir} = \langle{ U_{1}(z_{1},\overline{z}_{1})\cdots  U_{n}(z_{n},\overline{z}_{n})\rangle}.
\end{equation}
It is generally expected that there is some double copy structure relating the $S$-matrices of gauge theory and gravity. Here we see a natural manifestation of this in the form of the soft factors. Unlike standard double copy relations however, our prescription provides a direct map between the soft dynamics of the two theories by a simple squaring of a kinetic operator in the dual two-dimensional models. In summary, the double copy between the spin-$1$ and spin-$2$ soft $S$-matrices is obtained by the replacements,

\begin{equation}
    \begin{aligned}\partial\overline{\partial}&\rightarrow (\partial\overline{\partial})^{2}\\
e_{i}&\rightarrow \omega_{i}\kappa_{i}.\end{aligned} 
\end{equation}
It is worth noting that the phrase \emph{double copy} has only been used schematically. No relation is proposed here between this prescription and traditional double copies like BCJ or KLT. All we have indicated is that one moves from the spin-$1$ case to the spin-$2$ case by 'double copying', or squaring, a dynamical quantity, namely the kinetic operator\footnote{Thanks to Radu Roiban for comments that encoraged me to clarify this point.}.

The action (\emph{\ref{eq:3.5}}) determines a genuine two-dimensional model. Indeed, singularities of the sort exhibited by (\emph{\ref{eq:7}}) arise in the theory of crystal dislocations. Consider the tensor,

\begin{equation}
    \sigma_{ij} = \varepsilon_{ik}\varepsilon_{j\ell}\partial_{k}\partial_{\ell}\chi,
\end{equation}
where $\chi := \chi(x,y)$ depends on a source $\eta(x,y)$ through\footnote{Note also that we have switched to two-dimensional Cartesian coordinates.}

\begin{equation}
    \Delta^{2}\chi(x,y) = \eta(x,y)
\end{equation}
and the tensor $\varepsilon_{ij}$ is the Levi-Civita tensor in two dimensions. $\sigma_{ij}$ is the two-dimensional analogue of the stress tensor; it measures the stress and shear suffered by the system supporting dislocated lattice points. Given a suitably singular $\eta$ of the form

\begin{equation}
    \eta(x,y) = \sum_{a} b^{(a)}_{i}\partial_{i}\delta^{2}(x-x_{a},y-y_{a})
\end{equation}
the energy density

\begin{equation}
    \frac{1}{2}\int \chi(x,y)\eta(x,y)  d^{2}x
\end{equation}
characterizes that of a solid with a line defect normal to the plane charted by the $(x,y)$ coordinates. The vectors $b^{(a)}_{i}$ are known as \emph{Burgers} vectors, which characterize the direction of dislocation at each defect site. We note however that in our model the source function $\eta$ never quite becomes that singular, since we only need to consider sources of the form

\begin{equation}
    \eta(x,y) = \sum_{a} \kappa_{a}\delta^{2}(x-x_{a},y-y_{a}).
\end{equation}
This tells us that the theory we have on our hands is not of dislocations, which are line defects, but of objects known as disclinations \cite{PhysRevB.19.2457}, which experience a strongly confining interaction controlled by a biharmonic field equation and are point defects\footnote{I am much obliged to Suraj Shankar for pointing this out and explaining to me aspects of the theory of crystal defects.}.  We do not consider the theory of defects in larger detail here, reserving potential expansions of this point of view for future research. For interested readers however additional technical details on defects can be found in \cite{Kosterlitz_1973} and references therein.

With the results of this section and the last, we have determined that there exists a double copy construction that relates the soft $S$-matrix for photons to that of gravitons, albeit in a manner that is markedly different from the kinematical identities normally employed. Such a double copy structure is significant, since it appears to exist at the level of soft factors post integration over virtual momenta. Diagrammatically we have the summary

\begin{equation}\label{eq:20}
\begin{tikzcd}[column sep=70pt,row sep=50pt]
\mathcal{A}^{soft}_{n,s=1}|_{vir}\arrow[r,"\mathrm{Double\; Copy}"] \arrow[d,leftrightarrow,"\mathrm{Dual}"] & \mathcal{A}^{soft}_{n,s=2}|_{vir} \arrow[d,leftrightarrow,"\mathrm{Dual}"] \\%
\langle{V_1\cdots V_n\rangle} \arrow[r,"\Delta\longrightarrow \Delta^{2}"]& \langle{U_1\cdots U_n\rangle}.
\end{tikzcd}    
\end{equation}
In other words, a putative double copy is realised by first moving to the celestial description followed by squaring the kinetic operator.

\section{The Soft Theorems}
We have one more piece of the story to discuss, namely that of the soft theorems. So far, we have analyzed dualities between topological models that have been shown to reproduce the virtual soft $S$-matrices. The goal of this section is to establish that Ward identities in the two-dimensional model are responsible for the soft photon and graviton \emph{theorems}, arising due to the emission of real soft particles.

It is well known that the two-dimensional Coulomb gas model exhibits a so-called shift symmetry (see \cite{DiFrancesco:639405} for details). Specifically, the invariance of the action (\emph{\ref{eq:3}}) under global shifts of the form $\varphi \rightarrow \varphi + a$ implies the existence of a conserved holomorphic current, which takes the form,

\begin{equation}
    j_{s=1}(z,\overline{z}) = 8\pi\epsilon\overline{\partial}\varphi(z,\overline{z}).
\end{equation}
Incidentally, this can be derived by simple inspection of the equation of motion as well.

We can ask what would occur if we were to insert this current into the correlator (\emph{\ref{eq:6}})\footnote{I am indebted to Sudip Ghosh for suggesting this line of thought, which led to a complete rewriting of the present section.}. Doing so yields,

\begin{equation}
    \langle{j_{s=1}(z,\overline{z})V_1\cdots V_n\rangle} = \left(\frac{1}{\pi}\sum_{i}\frac{e_{i}}{z-z_{i}}\right)\langle{V_1\cdots V_n\rangle} 
\end{equation}
which may be readily verified. Indeed, this is precisely the soft photon theorem expressed in stereographic coordinates. It remains now only to verify the implication of the Ward identity, namely the fact that the integral

\begin{equation}
    Q_{s=1} = \oint_{\mathbb{CP}^{1}}j_{s=1}dz
\end{equation}
annihilates any correlator in which it is placed. Applying this to the foregoing equation we obtain,

\begin{equation}
    \left(\frac{2\pi i}{\pi}\sum_{i}e_{i}\right)\langle{V_1\cdots V_n\rangle} = 0.
\end{equation}
This condition, known as the neutrality condition in the CFT context tells us that total charge is conserved. This amounts to an alternative proof of gauge invariance as a consequence of the soft theorem due to our $2d$/$4d$ duality.

This analysis proceeds analogously for the spin-$2$ case. Here, the shift symmetry leads to the Noether current

\begin{equation}
    j_{s=2}(z,\overline{z}) = \overline{\partial}\Delta c(z,\overline{z}).
\end{equation}
An insertion thereof inside the correlator (\emph{\ref{eq:13}}) provides

\begin{equation}
    \langle{j_{s=1}(z,\overline{z})U_1\cdots U_n\rangle} = \left(\frac{1}{\pi}\sum_{i}\frac{\omega_{i}\kappa_{i}}{z-z_{i}}\right)\langle{U_1\cdots U_n\rangle} 
\end{equation}
which is none other than the soft theorem for the emission of a single soft graviton. The Ward identity indicating the annihilation of any correlator by the operator

\begin{equation}
    Q_{s=2} = \oint_{\mathbb{CP}^{1}}j_{s=2}dz
\end{equation}
reduces to the statement that

\begin{equation}
    \left(\frac{2\pi i}{\pi}\sum_{i}\omega_{i}\kappa_{i}\right)\langle{V_1\cdots V_n\rangle} = 0.
\end{equation}

For generic kinematics, the foregoing equation cannot be reconciled with energy conservation unless all the $\kappa_{i}$ are identified, which is a restatement of the equivalence principle due to Weinberg \cite{Weinberg:1964ew}.

For thoroughness, we provide one more proof of charge conservation and the equivalence principles as implications of invariance under shift symmetry. The line of reasoning follows that of \cite{DiFrancesco:639405}.

Let us consider a shift $\varphi\rightarrow \varphi + a$. Since this is a global symmetry, we demand that under it the correlator (\emph{\ref{eq:6}}) be invariant. However, performing this operation on (\emph{\ref{eq:6}}), we encounter a phase,

\begin{equation}
    e^{ia\left(e_{1}+\cdots +e_{n}\right)}.
\end{equation}
For this to equal unity for arbitrary $a$, we are led to require

\begin{equation}
    e_{1}+\cdots +e_{n} = 0
\end{equation}
which is the conservation of charge.

Applying this argument to the spin-$2$ case means that we study the effects of a shift $c \rightarrow c+ a$ on the correlator (\emph{\ref{eq:13}}). Indeed, we notice a familiar phase factor of

\begin{equation}
    e^{ia\left(\omega_{1}\kappa_{1}+\cdots +\omega_{n}\kappa_{n}\right)}.
\end{equation}
Insisting that this equal unity again forces us to set

\begin{equation}
    \omega_{1}\kappa_{1}+\cdots +\omega_{n}\kappa_{n} = 0,
\end{equation}
which is only commensurate with generic kinematics when the $\kappa_{i}$ are identical, again yielding the equivalence principle\footnote{If one writes the disclination Lagrangian in the covariant form $8\pi\epsilon \partial_{i}\partial_{j}c\partial^{i}\partial^{j}c$, the transformation $c(z,\z) \rightarrow c(z,\z) +a_1+a_2 z + a_3 \z + a_4 z\z$ keeps the action invariant up to total derivatives. The corresponding conservation laws are simply the conservation of four-momentum. I thank Shamik Banerjee for pointing this out.}.

We point out parenthetically that multiple insertions of the soft currents lead to spurious double poles, which do not arise in the conventional soft theorems. Indeed, a systematical dual description of multiple soft emissions can be developed, and will appear in a forthcoming work.

\section{Discussion}

In this short note we have studied a dual model of soft divergences in quantum field theory. We related the soft $S$-matrix for theories with massless spin-$1$ particles to a dual model of topological defects on the worldsheet, known variously as the Coulomb gas model or XY model. This construction made it possible to identify a double copy prescription that automatically yields the soft $S$-matrix for theories with gravitons. The corresponding dual model when recognized as a model of crystal dislocations completes a satisfying web of dualities, summarized in (\emph{\ref{eq:20}}).

We suggest some possible avenues for future work. There has been much recent interest in worldsheet \cite{Cachazo:2013gna,Cachazo:2013hca,Cachazo:2013iaa,Cachazo:2014nsa,Cachazo:2014xea,Cachazo:2015aol,Mason:2009sa,Geyer:2015jch,Geyer:2016wjx,Geyer:2018xwu} and geometric models \cite{Mizera:2017cqs,Mizera:2017rqa,Arkani-Hamed:2017mur,Jagadale:2019byr,Kalyanapuram:2019nnf,Kalyanapuram:2020vil,Kalyanapuram:2020tsr,Kalyanapuram:2020axt} of scattering amplitudes in quantum field theory. In particular, certain two dimensional models at null infinity have been shown to reproduce as correlation functions scattering amplitudes in gauge theory \cite{Adamo:2015fwa} and gravity \cite{Adamo:2014yya}. A synthesis of this work with the ideas in this paper may reveal a consistent picture of massless amplitudes at null infinity.

Since we have restricted our attention to Abelian theories with no interacting light matter, the complications of gauge theories have not troubled us in this note. While gauge theory soft amplitudes are corrected at higher loop order and the gravitational analogues are not, the double copy has been shown to hold at the level of Feynman diagrams \cite{Oxburgh:2012zr}. Having only dealt with abelian interactions in this note, it remains to be seen with future work how the picture developed here generalises to the non-abelian case as well.

Vertex operators in the Coulomb gas picture are conformal primaries. While we have defined vertex operators by analogy for the gas of disclinations, a simple interpretation of the state generated by the vertex operator is unclear. One may regard such operators as point sources of disclinations, as the correlation function (\emph{\ref{eq:13}}) generates the free energy of a gas of dislocations. Finding a more satisfying interpretation can be a topic of further study\footnote{I thank Radu Roiban for raising this question.}. 

Since the theories given here are defined in two dimensions in an intrinsic fashion, they may be helpful in constructing holographic duals of gauge theory and gravity and study the corresponding double copy structures - at least at leading order factorization. It was observed that the KLT relations persist at low multiplicity on the celestial sphere \cite{Kalyanapuram:2020aya}. Synthesizing this latter work with the ideas of this paper would be of interest.

Finally, a natural and concrete step forward would be understanding how higher-order soft theorems can be derived in this framework.


\section*{Acknowledgements} 
It is a pleasure to thank Jacob Bourjaily for persistent encouragement and going over the draft.  I have benefited due to insightful comments from Nima Arkani-Hamed, Alfredo Guevara, Raghav Govind Jha, Seyed Faroogh Moosavian, Monica Pate and especially Vasudev Shyam. I thank Eduardo Casali for bringing his earlier work on worldsheet models at null infinity and issues beyond the eikonal limit to my attention. I wish to thank in particular Shamik Banerjee for pointing out how four-momentum conservation was realized in this framework, Sudip Ghosh for his suggestion to look at shift symmetries in the two-dimensional models, Radu Roiban for raising a number of questions that significantly improved the draft upon clarification and Suraj Shankar for giving me an instructive conceptual overview of the theory of crystal defects. This project has been supported by an ERC Starting Grant (No. 757978) and a grant from the Villum Fonden (No. 15369).



\bibliographystyle{utphys}
\bibliography{v1.bib}

\end{document}